
\input phyzzx


\overfullrule=0pt

\def\Ziii{Z_2}
\def\etal{{\sl et al.}}
\def\PL{{\sl Phys. Lett.\ }}
\def\NP{{\sl Nucl. Phys.\ }}

\def\LamQCD{\hbox{$\Lambda_{QCD}$}}

\def\({\left\lbrack}           \def\){\right\rbrack}
\def\{{\left\lbrace}           \def\}{\right\rbrace}

\def\Fmn{F_{\mu\nu}}           \def\FMN{F^{\mu\nu}}
\def\Fls{F_{\lambda\sigma}}    
\def\Fnl{F_{\nu\lambda}}       
\def\Hmn{H_{\mu\nu}}           \def\HMN{H^{\mu\nu}}
\def\Hls{H_{\lambda\sigma}}    
\def\Gmn{G_{\mu\nu}}           \def\GMN{G^{\mu\nu}}
\def\Lmn{L_{\mu\nu}}           
\def\Rmn{R_{\mu\nu}}           \def\RMN{R^{\mu\nu}}
\def\gmn{g_{\mu\nu}}

\def\psibar{\bar\psi}
\def\dmu{\partial_\mu}         \def\dnu{\partial_\nu}
\def\dM{\partial^\mu}         \def\dN{\partial^\nu}
         \def\ds{\partial_\sigma}
\def\Al{A_\lambda}   \def\As{A_\sigma}   
\def\Gl{G_\lambda}   \def\Gs{G_\sigma}   
\def\Jn{J_\mu}   \def\Jn{J_\nu}
\def\Jl{J_\lambda}   \def\Js{J_\sigma}   \def\Jr{J_\rho}
\def\JL{J^\lambda}   \def\JS{J^\sigma}   
\def\JM{J^\mu}   \def\JN{J^\nu}
\def\Ll{L_\lambda}      
\def\Rl{R_\lambda}   \def\Rs{R_\sigma}   \def\Rr{R_\rho}
\def\slasha#1{\setbox0=\hbox{$#1$}#1\hskip-\wd0\hbox to\wd0{\hss\sl/\/\hss}}
\def\slashb#1{\setbox0=\hbox{$#1$}#1\hskip-\wd0\dimen0=5pt\advance
       \dimen0 by-\ht0\advance\dimen0 by\dp0\lower0.5\dimen0\hbox
         to\wd0{\hss\sl/\/\hss}}

       \def\Vsla{\slashb{V}}
       \def\Asla{\slasha{A}}
       
\def\dsla{\slasha{\partial}}
\def\g5{\gamma_5}    \def\gm{\gamma_\mu} \def\gn{\gamma_\nu}
\def\gM{\gamma^\mu} 
\def\emn{\epsilon_{\mu\nu}}    \def\eMN{\epsilon^{\mu\nu}}
\def\eMNS{\epsilon^{\mu\nu\sigma}}
\def\eMNLS{\epsilon^{\mu\nu\lambda\sigma}}
\def\emnls{\epsilon_{\mu\nu\lambda\sigma}}
\def\eMNLSR{\epsilon^{\mu\nu\lambda\sigma\rho}}
\def\eff{{\hbox{\tenpoint eff}\,}}
\def\Jmt{\tilde J_\mu}         
\def\Jb{\bar J} \def\Jbm{\bar J_\mu} \def\Jbn{\bar J_\nu}
\def\Jbl{\bar J_\lambda} \def\Jbs{\bar J_\sigma}
\def\JbM{\bar J^\mu} \def\JbN{\bar J^\nu}
\def\JbL{\bar J^\lambda} \def\JbS{\bar J^\sigma}
             
\def\tum{\tilde U^{-1}}        \def\tu{\tilde U}
                               \def\tjm{\tilde J_\mu}

\REF\EMC{The EMC Collaboration, J. Ashman \etal,
\PL {\bf B206}(1988)364;
\NP {\bf B328}(1989)1.}

\REF\SMC{The SMC Collaboration, B. Adeva \etal,
\PL {\bf B302}(1993)533.}

\REF\SLAC{The E142 Collaboration, P.L. Anthony \etal,
{\sl Phys. Rev. Lett.} {\bf 71}(1993)959.}

\REF\BJsr{For recent theoretical analysis of the data see
J. Ellis and M. Karliner,
{\sl Phys. Lett.} {\bf B313}(1993)131 and
J. Ellis and M. Karliner,
{\sl Spin Structure Functions};
Plenary talk at
PANIC '93, CERN-TH.7022/93, hep-ph/9310272,
and references therein.}

\REF\GM{A.~Manohar and H.~Georgi,
{\sl Nucl.\ Phys.} {\bf B234}(1984)189.}

\REF\Skyrme{T.~H.~R.~Skyrme, {\sl Proc.\ Roy.\ Soc.\ London}
{\bf B260}(1961)127.}

\REF\SM{E.~Witten, {\sl Nucl.\ Phys.} {\bf B223}(1983)422 and
433;
G.~Adkins, C.~Nappi, and E.~Witten,
{\sl Nucl.\ Phys.}\ B{\bf 228}(1983)552;
$SU(3)_f$ extension in E.~Guadagnini,
{\sl Nucl.\ Phys.}\ B{\bf 236}(1984)35;
P.~O.~Mazur, M.~A.~Nowak, and \hbox{M.~Prasza\l owicz},
{\sl Phys.\ Lett.}\ B{\bf 147}(1984)137.}

\REF\WittenLewes{E.~Witten in {\em Lewes Workshop Proc.;}
A.~Chodos {\em et al.}, Eds; Singapore, World Scientific, 1984.}

\REF\Kaplan{D.~B.~Kaplan,
{\sl Phys.\ Lett.}{\bf B235}(1990)163;
{\sl Nucl.\ Phys.}{\bf B351},(1991)137.}

\REF\staticQ{G.~Gomelski, M.~Karliner and S.~B.~Selipsky,
{\em Static properties of quark solitons},
BUHEP-92-17/TAUP 2034-93, hep-ph/9304217 (1993).}

\REF\EFHK{J.~Ellis, Y.~Frishman, A.~Hanany and M.~Karliner,
{\sl Nucl.\ Phys.}\ {\bf B382}(1992)189.}

\REF\AndBon{A.~A.~Andrianov and L.~Bonora,
{\sl Nucl.\ Phys.} {\bf B233}(1984)232.}

\REF\Balog{J.~Balog,
{\sl Phys.\ Lett.}  {\bf B149}(1984)197.}

\REF\Manes{J.~L.~Manes,
{\sl Nucl.\ Phys.} {\bf B250}(1985)369.}

\REF\BarZu{W.~A.~Bardeen and B.~Zumino,
{\sl Nucl.\ Phys.} {\bf B244}(1984)421.}

\REF\Fujikawa{K. Fujikawa {\sl Phys.\ Rev.} {\bf D21}(1980)2848;
erratum {\sl ibid.}, {\bf D22}(1980)1499.}

\REF\GaLew{J. Gasser and H. Leutwyler
{\sl Nucl. Phys.} {\bf B250}(1985)465.}

\REF\ERT{D.~Espriu, E.~de~Rafael and J.~Taron,
{\sl Nucl.\ Phys.} {\bf B345}(1990)22.}

\REF{\WittenBoson}{E.~Witten, {\sl Comm. Math. Phys}
{\bf 92}(1984)455.}

\REF\mex{Y. Frishman Quark trapping in a model field theory. Mexico City
1973. Berlin, Heidelberg, New York: Springer 1975}

\REF\Bardeen{W.~A.~Bardeen,
{\sl Phys.\ Rev.} {\bf 184}(1969)1848.}

\REF\AGG{L.~Alvarez-Gaume and P.~Ginsparg,
{\sl Ann.\ Phys.}\ {\bf 161}(1985)423.}

\REF\Veneziano{G. Veneziano,
{\sl CP and $U(1)$ problems, and their relationship in QCD};
in La Plagne 1983 Proceedings, {\sl Beyond The Standard Model},
pp. 355-367.}

\REF\PolWieg{A.~Polyakov and P.~B.~Wiegmann, \PL {\bf B131}(1983)121.}

\REF\DS{P.H. Damgaard and R. Sollacher,
CERN-TH-7073/93, Nov. 1993, submitted to Phys. Lett.}

\nopubblock
\line{\hfill WIS-93/110/Nov-PH}
\line{\hfill TAUP-2117-93}
\line{\hfill hep-ph/9311250}
\line{\hfill}
\title{Quark Solitons from effective action of QCD}
\author {
Yitzhak Frishman\foot{e-mail: fnfrishm@weizmann.bitnet}
and Amihay Hanany\foot{e-mail: ftami@weizmann.bitnet}
}
\address{ Department of Particle Physics   \break
          Weizmann Institute of Science \break
          76100 Rehovot Israel}
\author {Marek Karliner\foot{e-mail: marek@vm.tau.ac.il}}
\address{School of Physics and Astronomy\break
         Beverly and Raymond Sackler    \break
         Faculty of Exact Sciences      \break
         Ramat Aviv Tel-Aviv, 69987, Israel}
\abstract{We derive an effective low energy action for QCD in 4
dimensions. The low energy dynamics is described by chiral fields
transforming non-trivially under both color and flavor.
We use the method of anomaly integration from the QCD action. The
solitons of the theory have the quantum numbers of quarks.
They are expected to be the constituent quarks of hadrons.
In two dimensions our result is exact, namely the bosonic gauged action
of WZW.}
\endpage

\chapter{Effective action from anomaly integration}
The connection between the phen\-o\-men\-ologically successful
non-relativistic con\-stit\-uent-quark model (NRQM),
and QCD's fundamental degrees of freedom continues to be
one of the open and puzzling questions in particle
physics.
The constituent quarks which are the basic building blocks of the NRQM
have the same quantum numbers as the much lighter and
highly relativistic QCD current-quark fields.  In addition, recent
data   from   polarized   deep   inelastic   lepton-nucleon
scattering\refmark{\EMC-\SLAC} and the ensuing conclusions about
spin structure of the  nucleon\refmark{\BJsr} indicate that the
constituent quarks are composite objects with internal structure. The
emergence of constituent quarks as low-energy effective excitations of
QCD is clearly due to a nonperturbative mechanism.  Finding a
theoretical description of this  mechanism is an important and
interesting challenge.

Some steps in this direction have already been made.  A model
combining  some  of the  features  of  the chiral  quark\refmark{\GM}
and
skyrmion\refmark{\Skyrme-\WittenLewes} approaches has been constructed
by Kaplan\refmark{\Kaplan}.  In this model it is postulated that at
distances smaller than the confinement scale but large enough to allow
for nonperturbative phenomena the effective dynamics of QCD is
described by chiral dynamics of a bosonic field which takes values in
$U(N_c {\times} N_f)$.  This effective theory admits classical
soliton solutions. Assuming that they are stable and may be quantized
semiclassically, one then finds that these solitons are extended
objects  with  spin $1/2$,  and that they  belong  to the  fundamental
representation of color and flavor.  Their mass is of order \LamQCD\
and radius of order $1/\LamQCD$.\refmark{\staticQ}
 It is very tempting to identify them
as the constituent quarks. Thus the constituent quarks in this model
are ``skyrmions'' in color space.

This is a very attractive idea, but some crucial elements of the
puzzle are still missing, most importantly a derivation of the
effective dynamics from QCD.  However, recent work on QCD in 1+1
dimensions (QCD$_2$) provides evidence in support of this picture.
\refmark{\EFHK} Exact nonabelian bosonization allows the Lagrangian of
QCD$_2$ to be re-expressed in terms of a chiral field $U(x) \in
U(N_c{\times}N_f)$ and of the usual gauge field.  This bosonized
Lagrangian has topologically nontrivial static solutions that have the
quantum numbers of baryons and mesons, constructed out of constituent
quark and anti-quark solitons.

When $N_c$ solitons of this type are combined, a static,
finite-energy, color singlet solution is formed, corresponding to a
baryon.  Similarly, static meson solutions are formed out of a soliton
and an anti-soliton of different flavors.  The stability of the mesons
against annihilation is ensured by flavor conservation.  These results
can be viewed as a derivation of the constituent quark model in
QCD$_2$.  Thus the idea of constituent quarks as solitons of a
Lagrangian with colored chiral fields becomes exact in
\hbox{$D = 1{+}1$}.

In this paper we will provide a constructive scheme for deriving
from QCD the relevant effective Lagrangian. In two dimensions
the procedure will yield the exact bosonic action of QCD$_2$,
namely the gauged WZW Lagrangian. In four dimensions the
resulting action turns out to be a chiral Lagrangian whose
leading terms are very similar to the action conjectured
by Kaplan.

The  construction of  the  effective  chiral action for
bosonic  variables is  based on  integration of  the anomaly
equation.\refmark{\AndBon-\Manes}
 The  idea is  to construct  an action  which will
reproduce  the anomaly  under  chiral transformations.   One
begins with the  anomaly equation as the  starting point and
integrates it with respect to the chiral rotation parameter.
The resulting action  typically contains additional
terms which  are anomaly-free.
These terms are  analogous to an integration  constant in an
ordinary indefinite integral. Therefore their coefficients
are arbitrary. There are also additional terms in the form of
Chern-Simons action which are non-local.

In order to resolve this freedom, one may adopt a strategy
analogous  to computing  a definite  integral, in this case rotating the
action by a finite chiral rotation. This rotated action is vector
invariant by choice of scheme. When the change in external fields is
compensated by a change in the chiral phase, the rotated action is also
axial invariant. Next, taking the difference
of  the rotated  and unrotated  action,  one gets the
effective chiral  action for  bosonic fields. This
action includes a term which is non-local, namely the Wess-Zumino
term, which locally is a total derivative.

We will  see that in  2 dimensions this  procedure generates
the gauged WZW action, while  in 4 dimensions it generates a
new type  of low-energy  effective action for  chiral fields
transforming non-trivially under both color  and flavor.  We
will then  demonstrate that  baryons and constituent quarks
are the solitonic excitations of such an action.

In order  to carry out  this program,  we need to  deal with
several preliminary  issues.  First,  we have to  define the
scheme in which the anomaly equation will be discussed. Having chosen the
scheme, we also need to choose a suitable regularization prescription.
In a given scheme the choice of
the regularization cannot influence the results of an exact
calculation. As we shall see, however, in 4 dimensions the
effective action contains an infinite number of terms,
corresponding to expansion in terms of powers of external
momenta divided by the cutoff. If we truncate the expansion
at a finite number of terms and take a finite cutoff, the
result will be regularization dependent. In 2 dimensions, however, the
limit of infinite cutoff reproduces the exact bosonized action of WZW.

Given an anomaly equation  one  can redefine  the  axial
current and add a counter  term which is an axial  variation
of a local action. \refmark{\BarZu} This creates an ambiguity  in  the
definition  of the quantum axial  current.  So  we need  some
principles which  remove this  ambiguity.  The first  one is our choice
of the vector  scheme, we  want  to remain with a vector-conserving
theory.  The second one is  to keep the right relation between
the axial and the vector currents. In two dimensions the axial current
is the dual of the vector current $J^\mu_5=-\eMN J_\nu$. In 4 dimensions
no such simple relation exist.
We choose the heat kernel regularization.\refmark{\Fujikawa}
In 2 dimensions this procedure is well defined  and the  end
result  is a complete  bosonization  for   gauged  fermions.
Unfortunately in 4  dimensions this cannot be  done and the
kinetic  term is  proportional to  the cutoff  squared.  We
therefore  keep  the  UV  cutoff finite  and  look  for  its
physical meaning.  This  will be described later.  The
price we pay is that we remain with regularization dependent
terms  and a  low energy  expansion.   The value  of the  UV
cutoff will determine  the range of energies  for which this
approximation  is valid.   What is  remarkable is  that these
calculations   reproduce   very   accurately   the   numbers
calculated  for chiral  Lagrangians in the case  of flavor
symmetry.\refmark{\GaLew}
It is still mysterious why the heat-kernel regularization
fits the results of experiment\refmark{\Balog,\ERT}

The outline of the paper
is as follows. In sec.~2 we introduce the notation and
set up the general
formalism. In sec.~3 we apply the formalism to
derive the gauged
WZW action in 2 dimensions. In sec.~4 we derive
the low energy effective action for the chiral field in 4 dimensions. In
section 5 we discuss various
schemes of choosing the group in which the chiral field is embedded,
new classical solutions in the ``$U$-scheme",
conserved currents including the baryon number current,
and finally the Polyakov-Wiegmann formula
extended to 4 dimensions. Its low energy
approximation, up to dimension-4 operators, is given explicitly.
\chapter{General formulation}
In this section we set the notation and describe the general
procedure.   Those features  which  are special  to  2 or  4
dimensions will be discussed in the relevant sections.

The fermionic part of the QCD
Lagrangian with  external sources coupled to  vector and axial
currents, scalar and pseudoscalar densities
labeled $V_\mu, A_\mu,  S, P$
respectively, has the form
$$S_F=
\psibar(i\dsla
-\Vsla-\Asla\g5-S-i\g5 P
)\psi.\eqn\lagqcd$$
Where $V_\mu\equiv V_\mu^aT^a$,
$A_\mu\equiv A_\mu^aT^a$, $S\equiv S^aT^a$, $P\equiv P^aT^a$,
$T_a\in G$ are the generators of some group G, normalized through
$\Tr(T^aT^b)={1\over2}\delta^{ab}$.
In the generic case we
think of $G$ as the group $U(N_c\times N_f)$, where
$N_c$ and $N_f$ are the number of colors and flavors, respectively.
Other groups can also
be chosen, as will be discussed below.
$V$ will eventually include the dynamical gauge field.
For later convenience we introduce the chiral fields
$$\eqalign{L_\mu&=V_\mu-A_\mu,\cr
R_\mu&=V_\mu+A_\mu.}\eqn\leftright$$
At this stage we do not include a mass term,
as this breaks the chiral symmetry explicitly, on top of the
breaking due to anomaly.
The variation of $S$ and $P$, on the other hand, is
compensated by a rotation of $\psi$ (see below).

Since \lagqcd\ is quadratic in the fermion fields, the
fermions can be integrated out explicitly, yielding the determinant
$$ Z\(V,A,S,P\)=\int\(D\psi\)\(D\bar\psi\)e^{iS_F}\quad.
\eqn\detdef$$
At this point we have some freedom of choice.
The axial current cannot be conserved. We need to choose a scheme
of conservation. Here, since at the end $V$ will contain the
dynamical field for gluons, we choose $Z$ to be invariant under vector
gauge transformations,
$$\{\eqalign{&\delta V_\mu=D_\mu\omega=\partial_\mu\omega+i\(
 V_\mu,\omega\)\cr
 &\delta A_\mu=i\(A_\mu,\omega\)\cr
 &\delta S=i\(S,\omega\)\cr
 &\delta P=i\(P,\omega\)\cr
 &\delta L_\mu=\partial_\mu\omega+i\(L_\mu,\omega\)\cr
 &\delta R_\mu=\partial_\mu\omega+i\(R_\mu,\omega\)
 .}\right.\eqn\gauge$$
On the other hand,
the variation of the determinant under the axial transformation
$$\{\eqalign{&\delta V_\mu=i\(A_\mu,\lambda\)\cr
 &\delta A_\mu=D_\mu\lambda=\partial_\mu\lambda+i\(
 V_\mu,\lambda\)\cr
 &\delta S=-\{P,\lambda\}\cr
 &\delta P=\{S
,\lambda\}\cr
 &\delta L_\mu=-\partial_\mu\lambda-i\(L_\mu,\lambda\)\cr
 &\delta R_\mu=\partial_\mu\lambda+i\(R_\mu,\lambda\)
 ,}\right.\eqn\axial$$
is anomalous.
This non-invariance of $Z$ under axial transformation can be written
in compact form as
$$-i{\delta\log Z\over\delta\lambda^a}=-{\cal A}^a\(V,A,S,P\)\eqn\anom$$
where ${\cal A}$ is the anomaly.
We integrate this equation using the chain rule
$${\delta Z\over\delta\lambda}=-D_\mu{\delta Z\over\delta A_\mu}
-i\(A_\mu,{\delta Z\over\delta V_\mu}\)+\{S
,{\delta Z\over\delta P}\}-\{P,{\delta Z\over\delta S}\}\eqn\chain$$
to get an action $Z_1$ which obeys
$$-i{\delta\log Z_1\over\delta\lambda^a}=-{\cal A}^a\(V,A,S,P\).
\eqn\chainII$$
$Z$ can be written as a product $Z_1Z_2$, where $Z_1$ is not chiral
invariant and can be calculated exactly from the anomalies, and
$Z_2$ is chiral invariant and cannot be calculated via this procedure.
We now make a right rotation with a chiral field $U$,
$$\{\eqalign{&L^U_\mu=L_\mu\cr
 &R^U_\mu=U^{-1}R_\mu U-U^{-1}i\dmu U\cr
 &(S+iP)^U=(S
+iP)U
\cr
 &(S-iP)^U=U^{-1}(S
-iP)
\cr
 }\right.\eqn\rght$$
and denote $Z^U=Z\(V^U,A^U,S^U,P^U\)$.
Later on we will show that the vector invariance implies that such a
rotation can be taken without loss of generality. Note that $U$ can be
chosen in any subgroup of $U(N_c\times N_f)$. This point will be also
discussed later.

We arrive at the connection
$$S_{F}=
S_\eff-i\log \Ziii\eqn\Sqcd$$
$$S_\eff=-i\log Z_1\(V,A,S,P\)+i\log Z_1\(V^U,A^U,S^U,P^U\).\eqn\leff$$
Now let us observe the following.
First, $\Ziii$ contains important physical information in it. It
has, in principle, any power of the external sources, therefore it
contains non-trivial information on general correlation functions.
Second, $\Ziii$ does not depend on U. Therefore all the $U$ physics is in
$S_\eff$. We will argue later that $U$ describes the dynamics of quark
solitons or baryon solitons, therefore $S_\eff$ describes their complete
dynamics. Third, note that in 2 dimensions, since we get exact
bosonization, $\Ziii$ becomes trivial. In the next sections we will treat
$S_\eff$ only. Note that $\log Z^U$ is invariant under the transformation
$$\{\eqalign{&\Phi\rightarrow\Phi^{\tilde U}\cr
                  &U\rightarrow\tilde U^{-1}U,}\right.\eqn\zuinv$$
where $\Phi$ represents any of the fields $L_\mu$, $R_\mu$, $S$, $P$ in
\rght. In fact $\Phi^U$ itself is invariant.
Note however that applying the variation of $\Phi$ in $\log Z_1$
reproduces the anomaly.
$S_\eff$ is invariant under the vector
transformation
$$\{\eqalign{&R_\mu\rightarrow\tum R_\mu\tu-\tjm\cr
 &L_\mu\rightarrow\tum L_\mu\tu-\tjm\cr
 &S\pm iP\rightarrow\tum(S\pm iP)\tu\cr
 &U\rightarrow\tum U\tu
 .}\right.\eqn\vector$$
where $\Jmt=i\tum\dmu\tu$.

We now turn to explain why the right rotation is sufficient in \rght.
The generalization of \rght\ to arbitrary left and right phases is
$$\{\eqalign{&L^g_\mu=g_L^{-1}L_\mu g_L-g_L^{-1}i\dmu g_L\cr
 &R^g_\mu=g_R^{-1}R_\mu g_R-g_R^{-1}i\dmu g_R\cr
 &(S+iP)^g=g_L^{-1}(S
+iP)g_R
\cr
 &(S-iP)^g=g_R^{-1}(S
-iP)g_L
 }\right.\eqn\lftrght$$
where $g$ denotes $(g_L,g_R)$ together. The effective action with both
left and right rotations is
$$S_\eff\(V,A,S,P,g_L,g_R\)
=-i\log Z_1\(V,A,S,P\)+i\log Z_1\(R^g,L^g,S^g,P^g\).\eqn\leffrl$$
Vector invariance \vector\ implies
$$S_\eff\(V,A,S,P,g_L,g_R\)=S_\eff\(V,A,S,P,1,g_L^{-1}g_R\)$$
and we can define $U=g_L^{-1}g_R$ to obtain \leff.
\chapter{The procedure in 2 dimensions}
In this section we calculate the effective action in 2 dimensions.
In this case the limit of
UV cutoff going to infinity is well defined.
We use $\Tr(\g5\gm\gn)=2\emn$, $\epsilon_{01}=1$, $g_{00}=-g_{11}=1$.
The axial variation of the action in 2 dimensions is given by
$$-i{\delta\log Z\over\delta\lambda^a}={1\over4\pi}\(\eMN(\Fmn-2i
\(A_\mu,A_\nu\))^a+2\left(D_\mu A^\mu\right)^a-2\{S
,P\}^a\)\eqn\axvar$$
$\Fmn$ is given by
$$\Fmn=\dmu V_\nu-\dnu V_\mu+i\(V_\mu,V_\nu\)+i\(
A_\mu,A_\nu\)={1\over2}(\Lmn+\Rmn),\eqn\fstr$$
and we have introduced the left and right field strengths
$$\eqalign{\Lmn&=\dmu L_\nu-\dnu L_\mu+i\(L_\mu,L_\nu\)\cr
\Rmn&=\dmu R_\nu-\dnu R_\mu+i\(R_\mu,R_\nu\).}\eqn\LR$$
Also, $D_\mu A_\nu = \partial_\mu A_\nu + i \(V_\mu,A_\mu\)$.
Note the term $D_\mu A^\mu$ in \axvar, which originates from a
$g_{\mu\nu}$ part rather than $\emn$ part. This term can
be calculated by recognizing that for $A=0$ the anomaly is $-{1\over2\pi}
\eMN\Fmn$ and, for $A\not=0$, $F$ is replaced by the field strength of
the new vector field $V_\mu+\emn A^\nu$.

Using the chain rule eq. \chain\ we have
an action which has the variation \axvar\
$$\eqalign{-i\log Z_1=S_{CS}\(R\)-S_{CS}\(L\)
&+{1\over2\pi}\int\eMN\Tr(V_\mu A_\nu)-{1\over2\pi}\int\Tr A_\mu A^\mu\cr
&+{1\over2\pi}\int\Tr(S^2)
}\eqn\logz$$
Where $S_{CS}$ is the Chern-Simons action
(an extension to 2+1 dimensions)
$$S_{CS}\(A\)={1\over4\pi}\int\eMNS\Tr(
A_\mu\dnu A_\sigma+{2\over3}iA_\mu A_\nu A_\sigma)\eqn\CS$$
which has the variation
$$\delta S_{CS}\(R\)={1\over4\pi}\int\eMN\Tr((\delta R_\mu)R_
\nu)+{1\over4\pi}\int\eMNS\Tr(\Rmn(\delta R_\sigma))\eqn\varCS$$
and when using $\delta R_\mu=\partial_\mu\omega+i\(R_\mu,\omega\)$ the
variation is
$$\delta S_{CS}\(R\)={1\over4\pi}\int\eMN\Tr((\dmu R_\nu)
\omega).\eqn\varCSo$$
We use the Chern-Simons anomaly contribution
$$S_{CS}\(R^U\)-S_{CS}\(R\)=-{1\over4\pi}\int\eMN\Tr(\Jbm R_\nu)
+{i\over12\pi}\int\eMNS\Tr(J_\mu J_\nu\Js)\eqn\CSan$$
where $J_\mu=U^{-1}i\dmu U$ and $\Jbm=-UJ_\mu U^{-1}$,
to calculate the effective action \leff,
$$\eqalign{S_\eff&={1\over8\pi}\int\Tr(J_\mu J^\mu)
-{i\over12\pi}\int\eMNS\Tr(J_\mu J_\nu\Js)\cr
&+{1\over4\pi}\int(g^{\mu\nu}+\eMN)\Tr\(\Jbm R_\nu
  -L_\mu U^{-1}R_\nu U+L_\mu J_\nu+L_\mu R_\nu\)\cr
&-{1\over8\pi}\Tr\{\(U^{-1}(S-iP)\)^2+\((S+iP)U\)^2+2(P^2-S^2)\}\cr
}\eqn\lis$$
Note that it is $R_-$ that couples to the right fermion current, and
$L_+$ to the left one. As can be seen the
action \lis\ is the gauged WZW action of level 1.\refmark{\WittenBoson}
 We learn that by
anomaly integration, the effective action is an exact bosonization of
Dirac fermions. Hints for this were given in ref. $\(\right.$\mex$\left.
\)$. Note that when $P\rightarrow0$ and $S\rightarrow M$, we get a
$U^2+U^{-2}$ term, but not the expected $U+U^{-1}$ term which corresponds
to a fermionic mass term. This is because we included only effects that
come from the anomaly, through the Jacobian of the change of variables in
fermion sector, but not the explicit breaking coming from the mass term.
\chapter{The procedure in 4 dimensions}
We now turn to 4 dimensions. We follow
exactly the same procedure as in the 2 dimensional case but here we
encounter some difficulties. While the result of eq. \lis\ is an exact
bosonization of fermions coupled to vector and axial sources, its
4-dimensional analogue is an approximation, valid only at low energies.
We use an expansion in powers of the cutoff, assuming the higher
terms are suppressed by powers of the cutoff squared.
The scale at which the approximation breaks down is determined
by $\Lambda$, the ultraviolet cutoff in the regularization procedure.

Let us now go over to the derivation. The variation of the determinant
under the axial transformation \axial, including terms up to zeroth power
of the cutoff, is given by
\refmark{\Bardeen,\AndBon}
$$
\eqalign{
-i{\delta\log Z\over\delta\lambda^a} & = {1\over4\pi^2}
\biggl\lbrace
\eMNLS
\biggl[
\Fmn\Fls + {1\over3} \Hmn\Hls - {4\over3}
(A_\mu A_\nu\Fls+A_\mu\Fnl\As
\cr
+\Fmn\Al\As) & + {8\over3} A_\mu A_\nu\Al\As
\biggr]
+16\Lambda^2D_\mu
A^\mu+{2\over3}\(D^\mu\Fmn,A^\nu\)+{1\over3}\(\Fmn,D^\mu A^\nu\)
\cr
&+{1\over3}\{D^\mu A_\mu,A_\nu A^\nu\}+2A_\mu D_\nu A^\nu A^\mu+{1\over6}
\(\(A_\mu,A_\nu\),\HMN\)
\cr
+ {1\over3} (D_\mu & A_\nu A^\mu A^\nu+A_\mu A_\nu D^\nu A^\mu)
+ D_\mu A_\nu A^\nu A^\mu+A_\mu A_\nu D^\mu A^\nu
\biggr\rbrace
+{\cal O} (\Lambda^{-2})
\cr
}\eqn\Anofd$$
where $\Hmn={1\over2}(\Rmn-\Lmn)$.
In \Anofd\ we have terms which are contracted with the metric
rather than with the
anti-symmetric tensor. Their coefficients are determined from
a heat-kernel regularization procedure.\refmark{\AndBon,\Fujikawa}
We do not have here a relation like in
2 dimensions between vector and axial currents,
and therefore we do not see simple relations among the various
terms in \Anofd.
Using the chain rule \chain, an
action which has the above variation is given by
$$ \eqalign{-i\log Z_1&=-{2\Lambda^2\over\pi^2}\int\Tr(A_\mu A^\mu)
+S_{CS}^5\(R\)-S_{CS}^5\(L\)\cr
&+{i\over48\pi^2}\int\eMNLS\Tr\(i(\Rmn+\Lmn)\{\Ll,\Rs\}
+(L_\mu L_\nu-{1\over2}L_\mu R_\nu+R_\mu R_\nu)\Ll\Rs\)\cr
&-{1\over12\pi^2}\int\Tr\{-{1\over4}(\Fmn)^2+\FMN\(A_\mu,A_\nu\)
+{1\over2}(D_\mu A^\mu)^2-(A_\mu A^\mu)^2\}}\eqn\Acanfd $$
where $S_{CS}^5\(R\)$ is the Chern-Simons action in 5 dimensions
(see for example the review in Ref.~[\AGG].)
$$S_{CS}^5\(R\)={1\over24\pi^2}\int\eMNLSR\Tr(R_\mu\dnu\Rl\ds\Rr+
{3\over2}iR_\mu R_\nu\Rl\ds\Rr-{3\over5}R_\mu R_\nu\Rl\Rs\Rr).\eqn\CSf$$
After rotation by $U$ we get the final result eq. \leff,
$$\eqalign{S_\eff&={\Lambda^2\over2\pi^2}\int\Tr\(J_\mu J^\mu+2\Jbm
R^\mu-2L_\mu U^{-1}R^\mu U+2L_\mu J^\mu+2L_\mu R^\mu\)\cr
&-{1\over240\pi^2}\int\eMNLSR\Tr(J_\mu J_\nu\Jl\Js\Jr)\cr
&+{i\over48\pi^2}\int\eMNLS\Tr\biggl\lbrace i\Rmn\{\Rl,\Jbs\}-R_\mu
R_\nu\Rl\Jbs\cr
& + {3\over2}R_\mu\Jbn\Rl\Jbs+\Jbm\Jbn\Jbl\Rs\biggr\rbrace\cr
+{i\over48\pi^2}\int&\eMNLS\Tr\{i(\Rmn+\Lmn)\{\Ll,\Rs\}
+(L_\mu L_\nu-{1\over2}L_\mu R_\nu+R_\mu R_\nu)\Ll\Rs\}\cr
-{i\over48\pi^2}\int&\eMNLS\Tr\{i(U^{-1}\Rmn U+\Lmn)\{\Ll,\Rs^U\}
+(L_\mu L_\nu-{1\over2}L_\mu R^U_\nu+R^U_\mu R^U_\nu)\Ll\Rs^U\}\cr
&+{1\over192\pi^2}\int\Tr\(-2(D_\mu U^{-1}D^\mu U)^2+(D_\mu U^{-1}D_\nu
U)^2+2(D_\mu D^\mu U^{-1})D_\nu D^\nu U\right.\cr
&\left.+4(\Rmn D^\mu U^{-1}D^\nu U+\Lmn D^\mu UD^\nu U^{-1})+U^{-1}\Lmn
U\RMN-\Lmn\RMN\)\cr
&+16\{\FMN\(A_\mu,A_\nu\)+{1\over2}(D_\mu A^\mu)^2-(A_\mu A^\mu)^2\}\cr
}\eqn\ealr$$
where $D_\mu U=\dmu U+iR_\mu U-iUL_\mu$,
$D_\mu U^{-1}=\dmu U^{-1}+iL_\mu U^{-1}-iU^{-1}R_\mu$.
Since we are interested in effective action for QCD we set
$L_\mu=R_\mu=G_\mu$ in eq. \ealr, to obtain
$$\eqalign{S_\eff&={\Lambda^2\over2\pi^2}\int\Tr(D_\mu UD^\mu U^{-1})
-{1\over240\pi^2}\int\eMNLSR\Tr(J_\mu J_\nu\Jl\Js\Jr)\cr
&+{i\over48\pi^2}\int\eMNLS\Tr\{i\Gmn\(\Gl,\Jbs\)-(G_\mu G_\nu
-{3\over2}G_\mu\Jbn
+\Jbm\Jbn)\Gl\Jbs
\}\cr
-{i\over48\pi^2}\int\eMNLS&\Tr\(i(U^{-1}\Gmn U+\Gmn)\{\Gl,\Gs^U\}
+(G_\mu G_\nu-{1\over2}G_\mu G^U_\nu+G^U_\mu G^U_\nu)\Gl\Gs^U\)\cr
+{1\over192\pi^2}&\int\Tr\(-2(D_\mu U^{-1}D^\mu U)^2+(D_\mu U^{-1}D_\nu
U)^2+2(D_\mu D^\mu U^{-1})D_\nu D^\nu U\right.\cr
&\left.+4(\Gmn D^\mu U^{-1}D^\nu U+\Gmn D^\mu UD^\nu U^{-1})+U^{-1}\Gmn
U\GMN-\Gmn\GMN\)\cr}\eqn\eag$$
\chapter{Bosonization Schemes and Solutions}
\section{Different schemes}
In this section we introduce several schemes of the effective action.
In the previous section we derived the effective low-energy action for
a general coset $G/H$ with $H$ a subgroup of $G$.  The chiral phase
takes values in $G$, while the gauge field takes values in $H$. In
QCD, $H$ is $SU(N_c)$.  Depending on the physics one is
interested in, there are various choices for $G$. We list here those
which we consider the most important.
\pointbegin The ``$U$-scheme".
In this case we choose $U\in U(N_c\times N_f)$.
\point The ``product scheme".
The group $G$ is $SU(N_c)\times SU(N_f)\times
U(1)$. One can parameterize $U$ by $U=h \cdot g$,
where $h\in SU(N_c)$ and $g\in U(N_f)$.
\point The effective $U(1)$ action.
It is an effective action for the $\eta^\prime$.
It is related to $\det U$\refmark{\Veneziano}.
\section{Classical solutions in the $U$-scheme}
The classical soliton solutions are embedded in an $SU(2)$
subgroup of $U(N_cN_f)$. They
have the form
$$U_c=e^{if(r)\vec\tau\cdot\hat r}$$
where $\vec\tau$ are the generators of this $SU(2)$ subgroup and $f(r)$
is a radial shape function with boundary conditions $f(0)=\pi$,
$f(\infty)=0$. There are three choices for $\vec\tau$.
\pointbegin $\vec\tau=\vec\sigma_c\times E_f$, where $E_f$ is a matrix in
flavor space which satisfies the relation $E_f^2=E_f$.
This is a colored soliton. This kind of soliton
has been discussed in detail
by Kaplan.\refmark{\Kaplan} Its baryon number is $\Tr_f(E_f)$.
\point $\vec\tau=E_c\times\vec\sigma_f$. If $E_c$ is the identity,
this is a color-singlet flavored soliton.
 This is the familiar Skyrme type soliton \refmark{\Skyrme,\SM},
widely discussed in the literature. Its baryon number is $\Tr_c(E_c)$.
\point A mixed colored flavored soliton. From the general discussion
in Ref.~\Kaplan\ and from the calculations in
2 dimensions \refmark{\EFHK} we expect that this new type of soliton
corresponds to a constituent quark.

The solitons of the theory can be taken in color space using the
following ansatz,
$$ U^i_c=e^{i f(r) (\vec\tau_c\cdot\vec r) E^i_f}
\eqn\ta$$
where $\vec\tau_c$ is in an
$SU(2)$ subgroup of $SU(N_c)$ and $E^i_f$ is a
matrix in flavor space with 1 in the $(i,i)$ entry and zeroes elsewhere.
In the normalization convention in which a nucleon has baryon number
$N_c$, the baryon number of this soliton is 1 and its
flavor quantum number is $i$, i.e. it has the quantum
numbers of a quark.

It is important to keep in mind, however, that the classical solution
with the quantum numbers of a constituent quark has a finite
energy. This is in clear contrast with the expectation that there are
no finite energy colored states. This is a clear deficiency of the
present treatment, and it can be traced back to the fact that in the
anomaly integration approach, confinement has to be put in by hand,
as it comes from the terms which we do not treat.

One may also consider solutions which carry baryon number different
from 1. In view of the results in 2 dimensions\refmark{\EFHK} we
expect  to  find  color-singlet solutions  constructed  out  of
soliton--anti-soliton pair and solutions which are constructed from
$N_c$ solitons. It is natural to interpret such solutions as mesons
and baryons and their constituent solitons as constituent quarks.
\section{Conserved currents}
The right and left currents are given by
$${\delta S_\eff\over\delta R_\mu}={\Lambda^2\over\pi^2}\JbM-{i\over48
\pi^2}\eMNLS\Jbn\Jbl\Jbs-{1\over48\pi^2}(\{\Jbn\JbN,\JbM\}-\Jbn\JbM\JbN
)\eqn\rightcur$$
$${\delta S_\eff\over\delta L_\mu}={\Lambda^2\over\pi^2}J^\mu+{i\over48
\pi^2}\eMNLS J_\nu\Jl\Js-{1\over48\pi^2}(\{J_\nu J^\nu,J^\mu\}
-\Jn\JM\JN)\eqn\leftcur$$
and the baryon number current by the trace of their sum,
$$J^B_\mu={i\over24\pi^2}\eMNLS\Tr(\Jn\Jl\Js).\eqn\Barno$$
Eqs.~\rightcur\ and \leftcur\
can be conveniently written in a compact form
$${\delta S_\eff\over\delta R_\mu}=\Tr\({1-\g5\over2}\gM\slasha{\Jb}
({\Lambda^2\over2\pi^2}-{1\over96\pi^2}\slasha{\Jb}\slasha{\Jb})\)\eqn
\colfcur$$
$${\delta S_\eff\over\delta L_\mu}=\Tr\({1+\g5\over2}\gM\slasha{J}
({\Lambda^2\over2\pi^2}-{1\over96\pi^2}\slasha{J}\slasha{J})\)\eqn
\cortcur$$
\section{The Polyakov-Wiegmann formula\refmark{\PolWieg}}
We want to find the effective action for 2 chiral phases.
The basic identity follows from the definition of the effective action
$S_\eff$ \leff.
$$S_\eff\(L,R,U_1U_2\)=S_\eff\(L,R^{U_1},U_2\)+S_\eff\(L,R,U_1\)\eqn\PW$$
For zero sources we have
$$S_\eff\(0,0,U_1U_2\)=S_\eff\(0,0,U_1\)+S_\eff\(0,0,U_2\)+S_1\(U_1,U_2\)
$$
which defines $S_1\(U_1,U_2\)$ as
$$S_1\(U_1,U_2\)=S_\eff\(0,-J_1^\mu,U_2\)-S_\eff\(0,0,U_2\)\eqn\Sone$$
For 2 dimensions we have\refmark{\PolWieg}
$$S_\eff\(0,0,U\)={1\over8\pi}\int\Tr(J_\mu J^\mu)
-{i\over12\pi}\int\eMNS\Tr(J_\mu J_\nu\Js)\eqn\eqWZW$$
and
$$S_1=\(U_1,U_2\)=-{1\over4\pi}\int(\gmn+\emn)\Tr(J_2^\mu\JbN_1).\eqn
\Sonetd$$
For 4 dimensions we have
$$\eqalign{S_\eff\(0,0,U\)&={\Lambda^2\over2\pi^2}\int\Tr(J_\mu J^\mu)
-{1\over240\pi^2}\int\eMNLSR\Tr(J_\mu J_\nu\Jl\Js\Jr)\cr
&+{1\over192\pi^2}\int\Tr\(-2(J_\mu J^\mu)^2+(J_\mu J_\nu)^2
+2(\dmu\dM U^{-1})\dnu\dN U\)\cr}\eqn\Sezes$$
$$\eqalign{S_1\(U_1,U_2\)&=-{\Lambda^2\over\pi^2}\int\Tr\JbM_2
J_1^\mu+{1\over48\pi^2}\int\emnls\Tr(J_1^\mu J_1^\nu\JL_1\JbS_2\cr
&+{3\over2}J_1^\mu\JbN_2\JL_1\JbS_2
-\JbM_2\JbN_2\JbL_2\JS_1)\cr}\eqn\Sefffd$$
\chapter{Summary and outlook}
We have derived an effective low energy action for QCD in 4 dimensions.
We have shown that the solitons in the theory have quantum numbers of
quarks. We interpret these solitons as constituent quarks.
An interesting plan might be to calculate the equations of motions
numerically and compare the results to experimental values. Other
quantities such as the $g_A$ of the constituent quarks might be
calculated.
In this case $g_A$ has a classical contribution too,
in analogy with the Skyrme model.
Multi-soliton solutions can also be considered as baryons and mesons
constructed out of quark solitons.
\ack{
The authors would like to thank John Ellis for useful
discussions.
Y.F. would like to acknowledge interesting
discussions with W.~Bardeen.
A.H. Would like to thank Adam Schwimmer, Jacob Sonnenschein and
Yigal Shamir for useful discussions.

The research of Y.F. and A.H. was supported in part
by the Basic Research Foundation administered by the
Israel Academy of Sciences and Humanities.
The research of M.K.  was supported in part
by the Einstein Center at the Weizmann Institute,
by the Basic Research Foundation administered by the
Israel Academy of Sciences and Humanities
and by grant No.~90-00342 from the United States-Israel
Binational Science Foundation(BSF), Jerusalem, Israel.
} 

\bigskip\noindent
\undertext{Note Added:}
Upon the completion of this work we have received a draft
of the paper
by Damgaard and Sollacher\refmark{\DS} in which the effective
action in four dimensions is derived by closely related
techniques. We thank the authors for communicating their
results to us prior to publication.

\refout
\bye